\documentclass[aps,prl,preprint,superscriptaddress,showpacs]{revtex4-1}

\newcommand{\figI}   [1]{\ref{fig:east_diagnostic}#1} 
\newcommand{\figII}  [1]{\ref{fig:hesel_diagnostic}#1} 
\newcommand{\figIII} [1]{\ref{fig:Last_Cycle_fig_3}#1}
\newcommand{\figIIII}[1]{\ref{fig:Figure_4}#1} 
\newcommand{\unit}[1]{\ensuremath{\, \mathrm{#1}}}
\newcommand{\pfrac}[2]{\frac{\partial#1}{\partial#2}}
\newcommand{\ExB}{$\bm{E}\times\bm{B} \,$}

\usepackage{amsmath}    
\usepackage{bm}
\usepackage{graphicx}   
\usepackage{verbatim}   
\usepackage{color}      
\usepackage{subfigure}  

\begin{document}


\title{Simulation of transition dynamics to high confinement in fusion plasmas}

\author{A H. Nielsen}
\affiliation{Department of Physics, Technical University of Denmark, Fysikvej, DK-2800 Kgs.-Lyngby, Denmark}

\author{G.S. Xu}
\affiliation{Institute of Plasma Physics, Chinese Academy of Sciences, Hefei 230031, People's Republic of China}

\author{J. Madsen}
\author{V. Naulin}
\author{J. Juul Rasmussen}
\affiliation{Department of Physics, Technical University of Denmark, Fysikvej, DK-2800 Kgs.-Lyngby, Denmark}

\author{B.N. Wan}
\affiliation{Institute of Plasma Physics, Chinese Academy of Sciences, Hefei 230031, People's Republic of China}

\date{\today}

\begin{abstract}
The transition dynamics from the low (L) to the high (H) confinement mode in magnetically confined plasmas is investigated using a first-principles four-field fluid model. Numerical results are in close agreement with measurements from the Experimental Advanced Superconducting Tokamak - EAST. Particularly, the slow transition with an intermediate dithering phase is well reproduced by the numerical solutions. Additionally, the model reproduces the experimentally determined L-H transition power threshold scaling that the ion power threshold increases with increasing particle density. The results hold promise for developing predictive models of the transition, essential for understanding and optimizing future fusion power reactors.
\end{abstract}


\maketitle

An outstanding issue in magnetic fusion research is the understanding of the transition between the Low (L-mode) and High (H-mode) confinement mode.  Although the H-mode is routinely achieved in a multitude of magnetic confinement devices, since its first observation more than 30 years ago, the transition still lacks (full) theoretical explanation and predictive modelling. 
The L-  to H-mode transition represents a characteristic feature of many complex non-linear systems, where an abrupt transition between two states is encountered in response to a  variation of some "control" parameters. 
Examples are the transport bifurcations in continuum systems, the transition from dominating convective cell transport to  global circulation in Rayleigh-Benard convection\cite{Mori_Kramoto_1998} and in geophysics the formation of transport barriers by zonal flows, e.g., the earth's polar vortex\cite{Dritschel-McIntyre_2008, McIntyre_2014}.

In magnetically confined plasmas, the generation and sustainment of global flows is observed to be a key ingredient in the transition\cite{Wagner_etal_82, Wagner_2007}. The transition behaviour can vary from very abrupt transitions to slow transitions with intermediate phases\cite{Zohm_1994, Wagner_2007}. 
The L-mode is characterized by relatively flat pressure profiles and significant turbulent particle and energy transport across the Last Closed Flux Surface (LCFS) into the region of open field lines - the Scrape-Off Layer (SOL). The H-mode, on the other hand, is characterized by the so-called pedestal of elevated pressure just inside the LCFS and a weak quiescent transport into the SOL leading to improved confinement.
The H-mode is of essential importance the operation and success of ITER - the next generation international fusion experiment. To achieve the goal of ignition, ITER will rely on low power access to the H-mode.

Recent experiments with advanced diagnostics provide detailed spatially and temporally information about the L-I-H transition dynamics. Here, the  I-phase refers to the transition phase between the L- and H-mode, characterized by strong quasi-periodic bursts of plasma into the SOL (the I-phase is also referred to as the dithering phase or limit-cycle oscillations (LCO)); see, e.g., \cite{Xu_etal_PRL_2011, Xu_etal_14, Schmitz_etal_12, schmitz2014a, Cheng_etal_13, Kobayashi_etal_13, kobayashi2014a}. 
Concurrent with the improved experimental diagnostics, new modelling approaches have been developed for the simulations of the coupled Edge-SOL dynamics. Important ingredients on the way towards improved understanding of the SOL dynamics have been to abandon the distinction between fluctuations and profiles and the usage of flux driven systems\cite{Garcia_etal_jnm_2007,naulin_jnm_2007,sarazin_etal_2000}. 

It has long been known that  the L-H transition is connected to the build-up of Zonal Flows (ZF), suspected to be triggered by turbulent Reynolds Stress (RS) and finally being sustained as Mean Flows (MF) driven by the steepened ion pressure gradient. The interaction between these players is complex and in principle may contain many elements, from electromagnetic perturbations to three-dimensional effects including details of the geometry.  Parts of the transition dynamics have been reproduced by heuristic zero and more recently one-dimensional  predator-prey type modelling, with one or two feedback loops acting on disparate time-scales (fast and slow), see, e.g., \cite{Kim_Diamond_03, dam2013a, Miki_etal_2012}, but the quantitative connection to actual experimental parameters is absent in these models. With the results presented here, we can connect these disparate worlds of heuristic and first principles modelling and calibrate the lower dimensional models. 

The L-H transitions, with particular focus on the L-I-H transition, is modelled numerically using the first principal fluid model HESEL\cite{Nielsen_et_al_2013,paperinpreparation}. The results  are compared directly with experimental observations from the Experimental Advanced Superconducting Tokamak - EAST. The robustness of the L-H transition indicates a basic mechanism that we believe is represented in the interaction of the two-dimensional electrostatic turbulence with the self-consistently developing profiles, including the interaction between ion pressure and flow, specifically. Important is also the role of the Edge-SOL coupling anchoring initial gradients to the LCFS. The HESEL model includes these elements, and has enabled us to perform detailed studies of the L-I-H dynamics for parameters determined solely by experiment. 

HESEL is an energy conserving four-field model based on the Braginskii equations\cite{Braginskii} governing the dynamics of a quasi-neutral, simple plasma. It describes interchange-driven, low-frequency turbulence in a plane perpendicular to the magnetic field at the outboard midplane. In the limit of constant ion pressure the model reduces to the ESEL model, which has successfully modeled fluctuations and profiles in JET\cite{Fundamenski_etal_NF_2007}, MAST\cite{Militello_etal_PPCF_2013}, EAST\cite{Ning_etal_2013} and TCV\cite{garcia_etal_05}. The HESEL model includes the transition from the confined region to the region of open field lines (SOL) and the full development of the profiles across the last closed flux surface (LCFS).  The model is solved in a local slab geometry with the unit vector $\hat{\bm{z}}$ along   the inhomogeneous toroidal magnetic field. The inverse magnetic field strength is approximated by $B_0/B = 1 + a/R + (\rho_s /R)$, where $a$ and $R$ are the minor and major radii, respectively. $x$ is the radial coordinate in the local slab and $B_0$ is a characteristic magnetic field strength. In the Bohm-normalization the model equations read: 
\begin{widetext}
 \begin{align}
    	  \frac{d}{dt}n + n\mathcal{K}(\phi) - \mathcal{K}(p_e)
	    &= \nabla \cdot (n \bm{u}_{R})- \frac{n}{\tau_n} ,
			\label{eq:n_norm}\\    
      \frac{d^0}{dt}  \bm{w}  + \{\nabla \phi, \nabla p_i\} - \mathcal{K}(p_e + p_i)    
      &=  \eta \nabla^2 w - \frac{w}{\tau_w}     
     +\frac{enc_s}{L_{\|}}\bigg[1-\exp\big(\frac{-e \langle\phi\rangle }{\langle T_e\rangle} \big)\bigg]  
		= \Lambda_w ,
		\label{eq:w_norm}\\
     \frac{3}{2}\frac{d}{dt}p_e + \frac{5}{2}p_e\mathcal{K}(\phi) -  \frac{5}{2}\mathcal{K}\big(\frac{p_e^2}{n}\big)                         
      &= 
       \nabla \cdot \bigg( \chi_{e\perp}\nabla T_e\bigg)
       -\frac{5}{2}\nabla \cdot (p_e \bm{u}_{R})
      - \bm{u}_{R} \cdot \nabla p_i       
      - \frac{T_e}{\tau_{p_e}} ,
     \label{eq:electronpressure}\\
     \frac{3}{2}\frac{d}{dt}p_i 
     + \frac{5}{2}p_i\mathcal{K}(\phi) 
     + \frac{5}{2}\mathcal{K}\big(\frac{p_i^2}{n}\big)  
     - p_i  \mathcal{K}(p_e + p_i)    
     &= 
      \nabla \cdot \bigg(\chi_{i\perp}\nabla T_i\ \bigg)
     -\frac{5}{2}\nabla \cdot (p_i \bm{u}_R)     
     + \bm{u}_{R} \cdot \nabla p_i             
	     - \frac{p_i}{\tau_{p_i}} 
	   + p_i\Lambda_w ,
     \label{eq:ionpressure}
 \end{align}
\end{widetext}
where $n$ is particle density, $\bm{w} = \nabla^2\phi + \nabla^2 p_i$ is the generalized vorticity, $\phi$ is the electrostatic potential, and $p_e$ and $p_i$ are electron and ion pressure, respectively. Temperatures are defined by $T_{i,e} = p_{i,e}/n$. Material derivatives and the magnetic field curvature operator are defined as
$\frac{d}{dt} = \pfrac{}{t} + \frac{1}{B} \hat{\bm{z}}\times \nabla \phi \cdot \nabla$,  
$\mathcal{K} = \nabla \big(\frac{1}{B}\big) \cdot \hat{\bm{z}} \times \nabla$,
except in the generalized vorticity equation where the material derivative $\frac{d^0}{dt}$ is taken with a constant magnetic field. Friction forces enter through the drift velocity $\bm{u}_{R} =- D(1+T_i/T_e) \nabla \ln n$, where $D$ is the neo-classical Pfirsch-Schl\"{u}ter diffusion coefficient\cite{Fundamenski_etal_NF_2007}. $\eta$ denotes the neo-classical viscosity coefficient, and $\chi_{e}$ and $\chi_{i}$ are the neo-classical, perpendicular electron and ion heat conduction coefficients, respectively. Losses due to advection along magnetic field lines in the SOL region is represented by the damping rates $\tau_n$, $\tau_w$, and $\tau_{p_i}$. Parallel electron heat conduction in the SOL region is parametrized by the damping rate $\tau_{p_e}$. The effect of sheath currents at material surfaces on which magnetic field lines in the SOL region terminate is approximated by an effective sheath dissipation term entering the vorticity Eq.~(\ref{eq:w_norm}), where $L_{\|}$ is the connection length and $\langle \cdot \rangle$ denotes a combined time and poloidal average. In the inner part of the closed field line region the fluid fields are forced towards profiles reflecting toroidal equilibrium.    

The generalized vorticity is the manifestation of the polarization current in the model and describes charge separation due to the inertia in the ion response to changes in the \ExB and diamagnetic drifts. The generalized vorticity appears to be responsible for driving the MF related to the pressure gradient and essential for setting up the edge transport barrier supporting the pressure pedestal in the H-mode. 

 We have applied the code to simulate experimental observations from the L-I-H transition campaign at EAST in 2012. These experiments employed the Gass Puff Images (GPI) diagnostics, which images, in 2D, the edge plasma turbulence in the plane perpendicular to the local magnetic field by looking at the emission from excited neutrals, the HeI-line. The intensity of the emission is related to the electron pressure, see, e.g., \cite{Maqueda_etal_NF_2003}. Using high speed cameras and correlation techniques it is further possible to derive quantities like velocities from the propagation of perturbations. A detailed description of the dual GPI system on EAST is found in Liu {\it et al}\cite{Liu_etal_RSI_2012}.  

The collision and parallel damping rate coefficients for all simulations presented in this paper are calculated using parameters characteristic for the plasma at the LCFS in EAST shot 41362:
$T_{e0} = 20 \unit{eV}$, $B_0 = 2.0 \unit{T}$, $q = 4$, $R = 2.0 \unit{m}$, $a = 0.5\unit{m}$, and distance from the LCFS to the limiter shadow $\Delta_{SOL} = 2.4\unit{cm}$. In the simulations the non-constant power input in EAST is emulated by ramping-up the ion temperature of the prescribed profile in the vicinity of the inner boundary as: $T_i = T_{i0} + (T_{i\mathrm{max}}-T_{i0}) \sin(\frac{\pi}{2}\frac{t}{t_{\text{ramp}}})$, where $T_{i0} = 20 \unit{eV}$ and $T_{i\mathrm{max}} = 60 \unit{eV}$.
The ion temperature dependent diffusion and parallel loss coefficients were adjusted accordingly.

Figures\,\figI{}-\figII{} display the time evolution of radial profiles and two integrated quantities from EAST shot 41362 and HESEL, respectively. The profiles are obtained by averaging over the poloidal simulation/measurement volume. In the simulation $n_0 = 1.5 \times 10^{19}\unit{m^{-3}}$ and $t_{\text{ramp}} = 20.88 \unit{ms}$. The experimental results are mostly from the GPI system\cite{Xu_etal_14}. 
The L-, I-, and H-phases are clearly recognized in the plots. 
Figure\,\figI{a} shows the evolution of the HeI line intensity profile, S,
Fig.\,\figI{b} shows the fluctuation part of S, and  Fig.\,\figI{c} shows the relative difference in the intensity S between radial positions $-0.7$ cm  and $1.5$ cm. 
During a plasma burst the intensity profile in Fig.\,\figI{a}  flattens significantly consistent with the breakdown of the pressure gradient observed in Fig.\,\figI{c}. 

\begin{figure}
 \includegraphics[width=1.0\columnwidth]{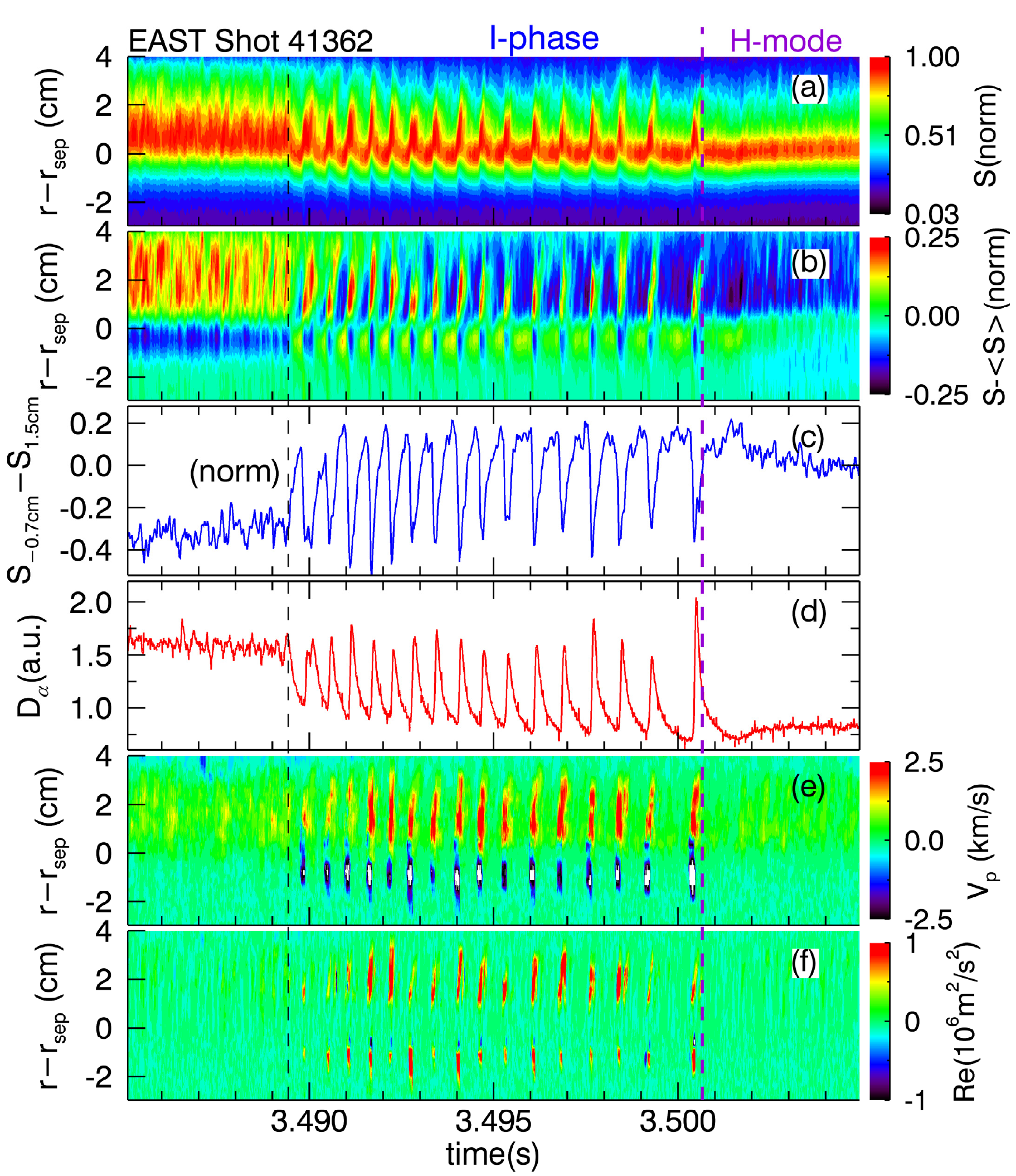}
\caption{EAST shot no. 41362 with input power slowly ramped up, revealing the essential features of the L-I-H transition.  The frames show measurements from the confined region $(r < 0)$ and the region with open field lines - the SOL $(r > 0)$.
a) The emission intensity of the HeI line, S,
b) the fluctuation level of S,
c) the difference in the relative GPI emission intensity between radial positions $-7$ mm  and $15$ mm,
d) the $D_{\alpha}$ emission from the outer divertor region,-
e) poloidal flow velocity from GPI,
f) turbulence-driven Reynolds stress: $<\tilde{v}_r \tilde{v}_p>$, the velocity fluctuations derived from GPI.}
\label{fig:east_diagnostic}
\end{figure}

The $D_{\alpha}$ signal, Fig.\,\figI{d}, originates from D atoms emitting in the divertor region. It is a measure for the amount of hot plasma there, which originates from perpendicular turbulent transport processes at the midplane into the SOL and subsequent parallel flow to the divertor.  $D_{\alpha}$ shows a similar evolution as the GPI intensity in the I-phase, but is lagging the fluctuations in the other signals consistent with the SOL parallel transport time from the out-board mid-plane to the divertor. 

The evolution of the poloidal velocity profile in Fig.\,\figI{e} and the corresponding Reynolds stress (RS) profile, Fig.\,\figI{f}, show I-phase oscillations, while in the H-mode velocity-fluctuations and RS signals are significantly reduced.  
We note, that measurements of velocity fields, by means of correlation techniques applied to propagating perturbations visible in the GPI signals, will fail in H-mode as the perturbation in the signal are too weak to correlate upon, see Fig.\,\figI{b}.

For comparing the experimental results to numerical simulations, we have generated a synthetic GPI diagnostics with "emission intensity" obtained from: $S_n \propto n_e n_n f(n_e,T_e)$, see Fig.\,\figII{a}, b, c. Here $f$ is the coefficient for excitation of helium and $n_n$ is the localized neutral gas density profile calculated at each point in time from neutral particles penetrating from the outer SOL and getting depleted by ionisation. 
We plot in Fig.\,\figII{d} the evolution of the integrated parallel particle flux at the outboard mid-plane as a proxy for the $D_\alpha$ signal in Fig.\,\figI{d}. By construction it lacks the time delay seen in the experimental data. 
The poloidal velocity profile, Fig.\,\figII{e},  and the corresponding Reynolds Stress (RS), Fig.\,\figII{f}, are directly obtained from the $E{\times}B$ velocities in the simulation. 
The plasma profile in H-mode is sustained by a substantial poloidal flow,  not detectable by the experimental GPI as mentioned above.  
In Fig.\,\figII{g} we show the evolution of the density profile, which follows closely the GPI intensity profile dynamics, besides the folding with the localised neutral gas profile. 

Comparing Figs.\,\figI{} and \, \figII{}  we observe a striking similarity between the experimental observations and the numerical results. The \textbf{I-phase} appears as a series of quiet periods, interrupted by large bursts of plasma reaching far into the SOL. The duration of the I-phase as well as the number of bursts are accurately matched between simulations and experiments. Note also that the period of the bursts are increasing as we ultimately approach the H-mode.

\begin{figure}
   \includegraphics[width=1.0\columnwidth]{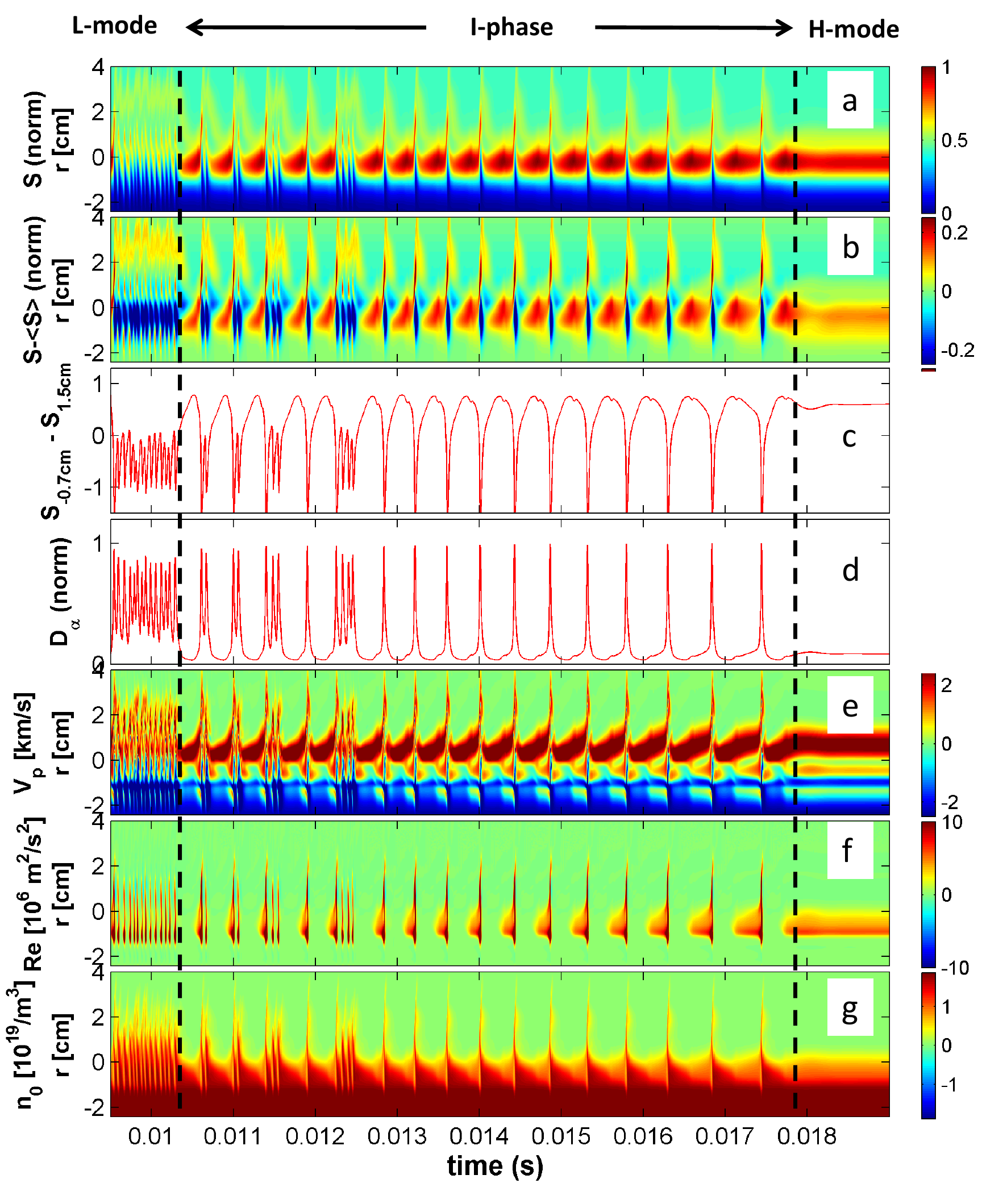}
\caption{\label{fig:hesel_diagnostic}
HESEL simulation for plasma parameters of EAST shot no 41362 similar to Fig.\, \figI.
a) to c) derived from synthetic GPI signal,
d) integrated parallel particle loss term as a proxy to divertor $D_\alpha$,
e) and f) derived from $E{\times}B$ velocities,
g) displays the electron density profile not accessible from GPI.}
\end{figure}

In Fig.\,\figIII{} we have shown the detailed evolution during the the last dithering cycle from the numerical simulations in Fig.\,\figII{}. We note that this cycle in overall appearance is similar to the other dithering cycles, except that this cycle makes the transition to H-mode. We observe three distinct phases. The first  is quiescent with nearly no fluctuations. It lasts until $T \sim 0.01742$ s, where significant fluctuations grow up and are dominating the dynamics. The pressure gradient is expelled from the edge deep into the SOL. 
In the final phase, from $T\sim 0.01746$ s the fluctuations and the enhanced transport into the SOL die out. The SOL is depleted of plasma and finally the pressure gradient across the LCFS re-establishes. 
The disappearance of the turbulent fluctuations is accompanied by the generation of a sheared zonal flow (ZF) through the RS. A strongly enhanced RS and significant Reynolds work coincide with the termination of the burst, where a poloidal flow is rapidly re-established. 
The generated poloidal flow, Fig.\,\figIII{c}, is later in parts sustained by the ion pressure gradient, Fig.\,\figIII{d}, induced mean flow (MF), but will decay gradually.
In the decay phase the RS, Fig.\,\figIII{e}, slowly increases again, 
but if the MF is not strong enough to sustain the barrier, it will lead towards the next burst and closing the cycle.
If, on the other hand, the MF, grows strong enough, this burst never occurs and a stable high confinement state is entered.
Thus, the dithering cycles appears at sub critical heat fluxes. As the input power increases the length of the dithering cycles increases in line with building up the ion pressure gradient and mean flow, until the system is able to enter the H-mode. 
The quiet phases during the I-phase thus resemble the H-mode, with the MF not yet strong enough to fully sustain it. 

\begin{figure}
   \includegraphics[width=1.0\columnwidth]{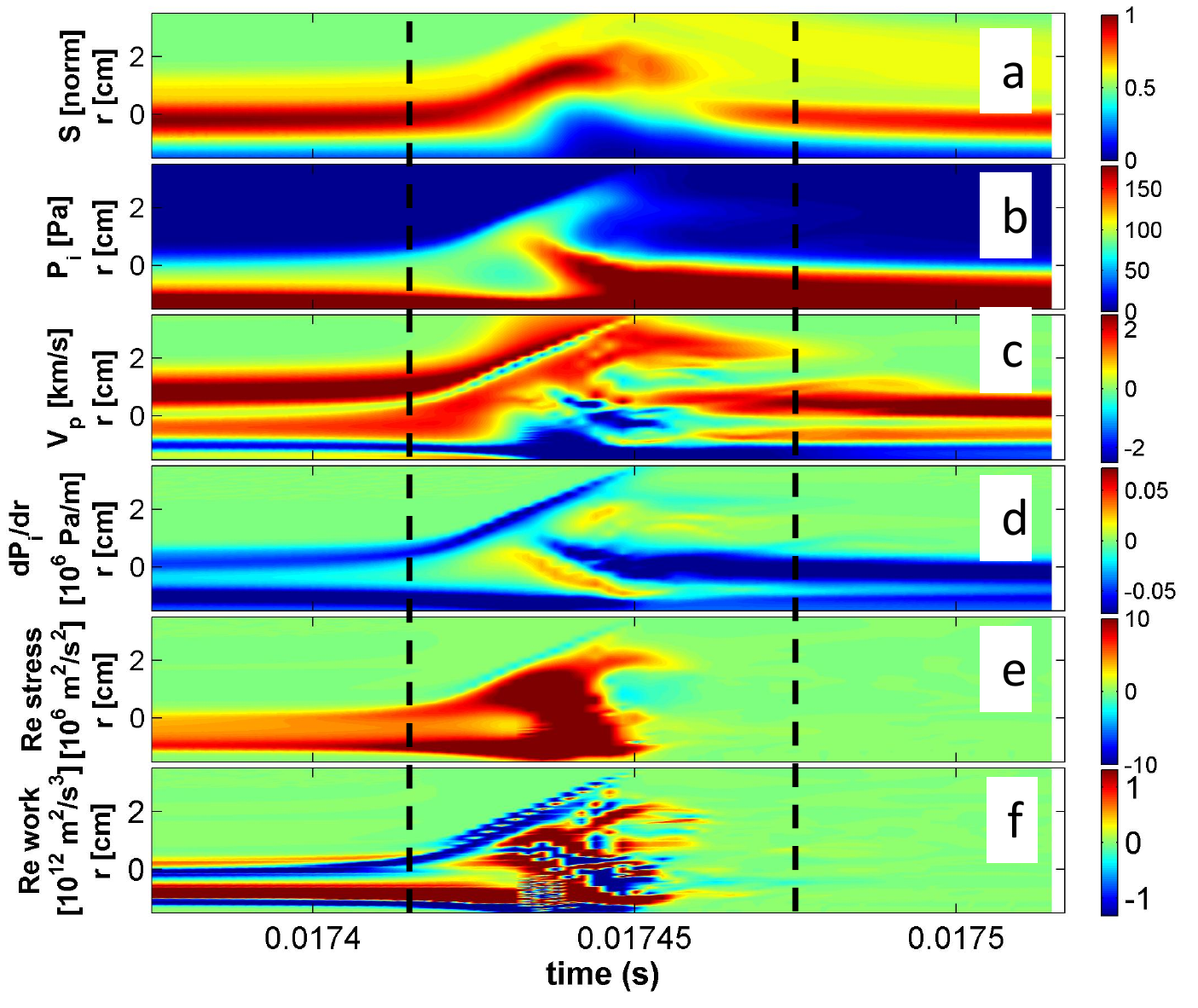}
\caption{\label{fig:Last_Cycle_fig_3}
Last cycle of the I-phase from HESEL. 
a) Synthetic GPI intensity, 
b) ion pressure,
c) poloidal velocity, 
d) ion pressure gradient, 
e) Reynolds stress,
f) Reynolds work: $<\tilde{v}_r \tilde{v}_p>\partial v_p /\partial r$.}
\end{figure}


The L-H-like transitions observed in the simulations are robust and have been obtained for a broad range of parameters. Different types of transitions have been observed ranging from abrupt transitions to slow transitions with an intermediate I-phase. In agreement with experimental observations\cite{Xu_etal_14} L-I-L transitions have also been observed when the power input is low. Figure\,\figIIII{a} shows an L-H transition without an I-phase, using a faster ramp-up of the of the ion temperature, $t_{\text{ramp}} = 2.61 \unit{ms}$. As the ion temperature is ramp-down a clear hysteresis is observed with a H-L transition occurring for an ion pressure reaches a value corresponding to a decrease in ion temperature at constant density by approximately $10 \unit{eV}$ compared to the L-H threshold. After the H-L transition the energy flux reverts to the same level as before the transition, a behaviour routinely observed in experiments\cite{Wagner_2007}. 
Assuming that the heat flux crossing the LCFS is axisymmetric distributed and originates from a region $30$ degree poloidally above and below the outboard midplane the total heat flux to material surfaces, neglecting radiation, can be estimated from the HESEL simulations to be $1.2$ MW for the LH transition, in close agreement to the estimated experimental input power of $1.0$ MW for this particular shot, \cite{Xu_etal_14}.  

\begin{figure}
\begin{tabular}{cc}
   \includegraphics[width=0.5\columnwidth]{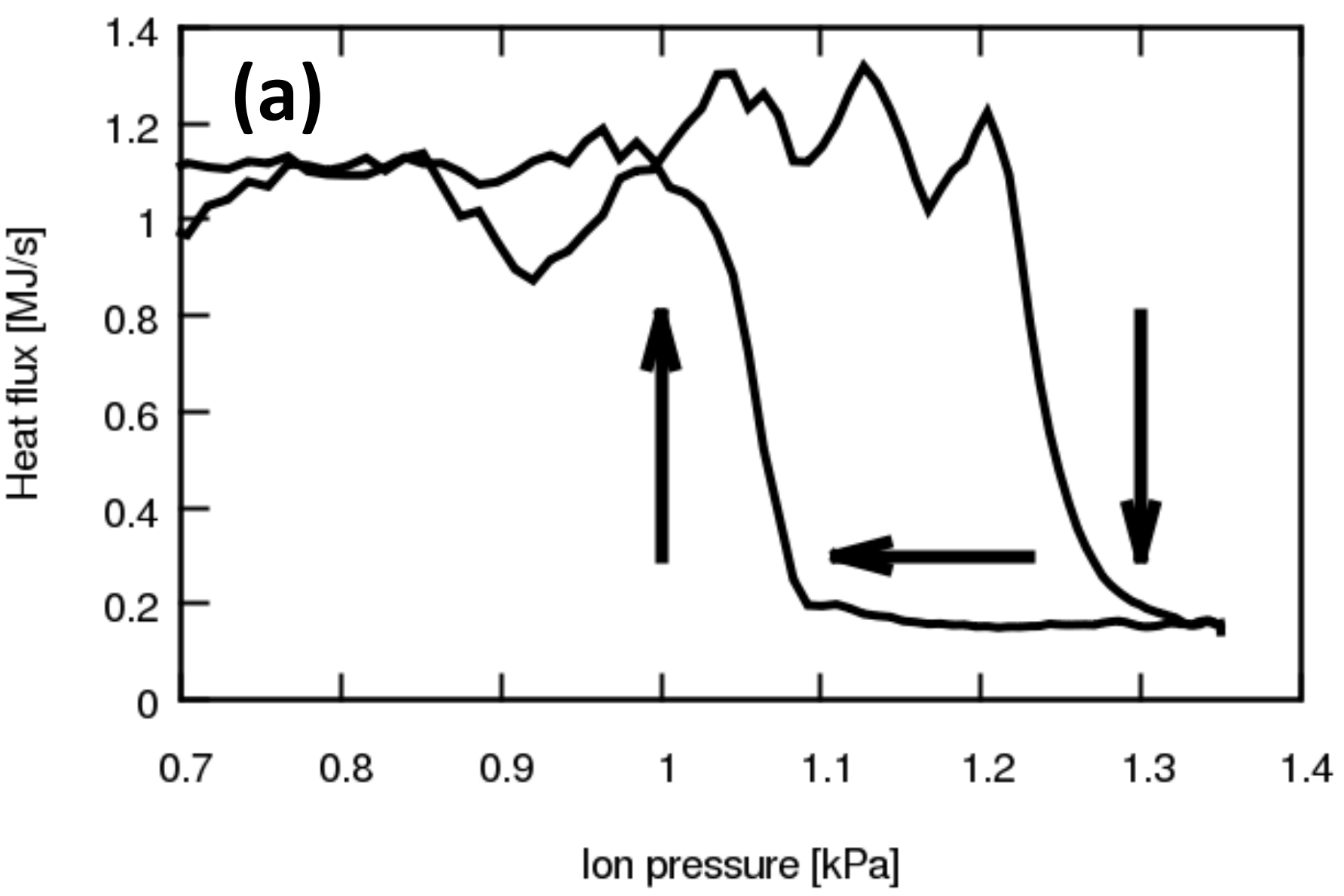}&
   \includegraphics[width=0.5\columnwidth]{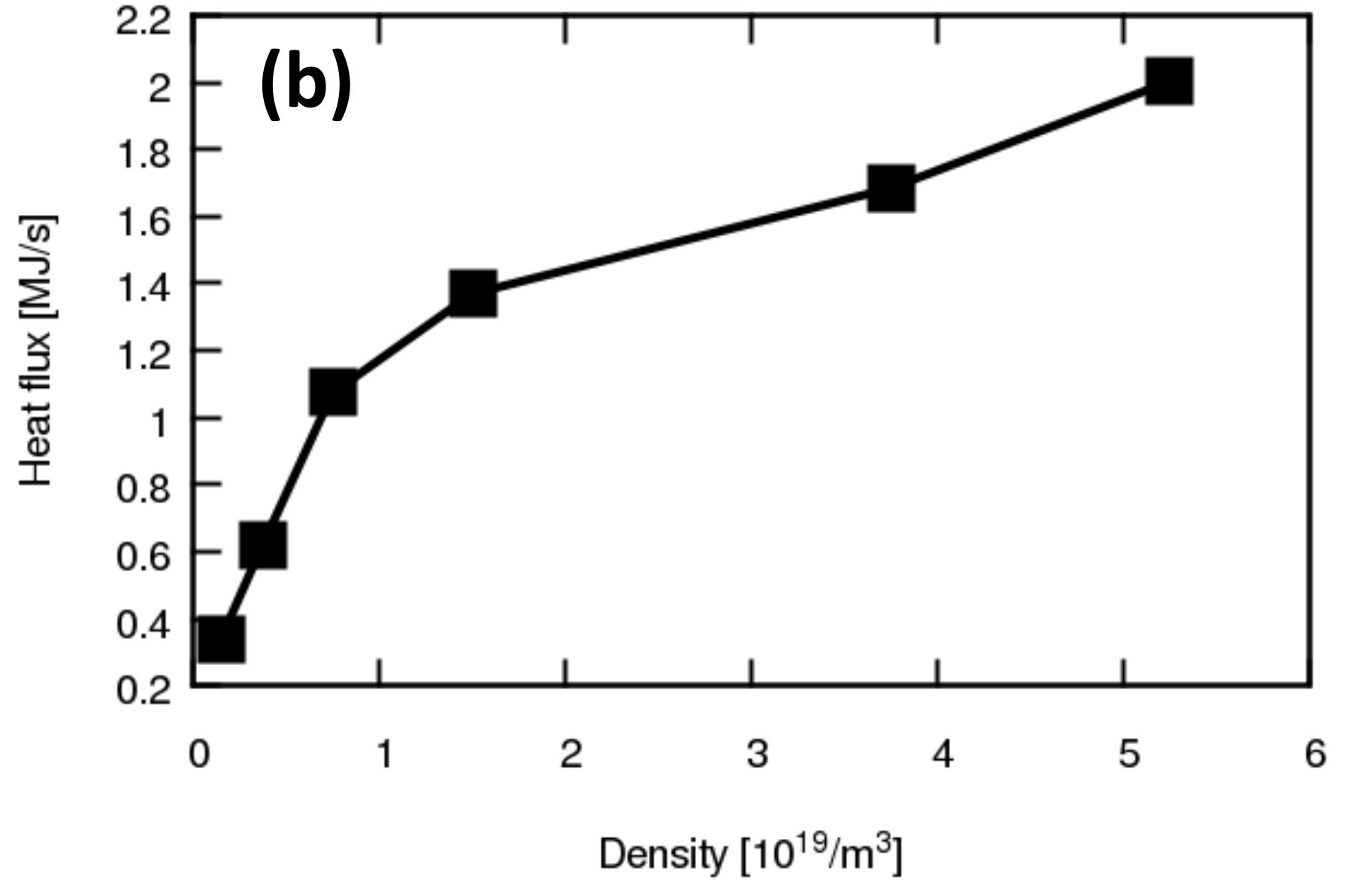}
	\end{tabular}
\caption{\label{fig:Figure_4}
HESEL simulations displaying a) the total heat flux across the LCFS as a function of the ion pressure at the LCFS during a transition, b)  heat flux across the LCFS at the transition for different densities.}
\end{figure}

The simulations also give insight into the general scaling of the L-H transition threshold power.
Figure \,\figIIII{b} show a series of HESEL simulations where only the reference particle density $n_{0}$ was varied. We observe that the transition threshold power increases at increasing particle density, 
as observed in the high density branch of the L-H transition threshold power\cite{Wagner_2007}. Commonly observed is also a fast increase of the L-H power threshold at lower densities, leading to a well-defined density with minimal power necessary to enter H-mode. 
In these experiments electrons are dominantly heated centrally. However, it was recently demonstrated that the ion heat channel  plays the key role in the L-H  transition\cite{Ryter_etal_2014}. Therefore, the ramp-up of the ion energy flux across the LCFS depends on the energy coupling between electrons and ions over the whole plasma volume.
When the power input is directly through the ions one should not expect to find a roll-over of the threshold power at a specific density. Our simulation results agree with this observation. 

To the best of our knowledge this is the first modelling of the L-H transition based on a first principle model reproducing vast details of the transition behaviour without free parameters. This still much simplified model appears to include the necessary and essential ingredients for the transition behaviour, but is still not a fully predictive model.
The results presented here form an essential step connecting the zero- and one-dimensional heuristic transition models with a predictive model and thus the full set of toroidal plasma dynamics. The ITER experiment, which relies on controlled H-mode access, needs this gap in understanding to be bridged.

\begin{acknowledgments}
This work was supported by the National Magnetic Confinement Fusion Science Program of China under Contracts No. 2011GB107001. JM was supported by an EFDA fusion researcher fellowship (WP11-FRF-RISOE/MADSEN).
\end{acknowledgments}


\end{document}